\documentclass[prb,twocolumn,preprintnumbers,amsmath,amssymb,superscriptaddress]{revtex4}
\usepackage[dvips]{graphicx}% Include figure files
\usepackage{dcolumn}% Align table columns on decimal point
\usepackage{bm}% bold math
\usepackage{longtable}
\usepackage[dvips]{color}
\usepackage{ulem}

\begin{document}

\preprint{APS/123-QED}

\title{Detailed dynamics of a moving magnetic skyrmion lattice in MnSi observed using a small-angle neutron scattering 
under an alternating electric current flow}
% Force line breaks with \\

\author{D. Okuyama}%
\email[]{okudaisu@post.kek.jp}
\affiliation{Institute of Multidisciplinary Research for Advanced Materials (IMRAM), Tohoku University, Katahira 2-1-1, Sendai 980-8577, Japan.}
\affiliation{Institute of Materials Structure Science (IMSS), High Energy Accelerator Research Organization (KEK), Oho 1-1, Tsukuba, Ibaraki 305-0801, Japan}
\author{M. Bleuel}%
\affiliation{NIST Center for Neutron Research, National Institute of Standards and Technology, 100 Bureau Drive, Gaithersburg, Maryland 20899-8562, USA}
\affiliation{Department of Materials Science and Engineering, University of Maryland, College Park, MD 20742-2115, USA}
\author{Q. Ye}%
\affiliation{NIST Center for Neutron Research, National Institute of Standards and Technology, 100 Bureau Drive, Gaithersburg, Maryland 20899-8562, USA}
\affiliation{Department of Materials Science and Engineering, University of Maryland, College Park, MD 20742-2115, USA}
\author{J. Krzywon}%
\affiliation{NIST Center for Neutron Research, National Institute of Standards and Technology, 100 Bureau Drive, Gaithersburg, Maryland 20899-8562, USA}
\author{N. Nagaosa}%
\affiliation{RIKEN Center for Emergent Matter Science (CEMS), Wako 351-0198, Japan}
\author{A. Kikkawa}%
\affiliation{RIKEN Center for Emergent Matter Science (CEMS), Wako 351-0198, Japan}
\author{Y. Taguchi}%
\affiliation{RIKEN Center for Emergent Matter Science (CEMS), Wako 351-0198, Japan}
\author{Y. Tokura}%
\affiliation{RIKEN Center for Emergent Matter Science (CEMS), Wako 351-0198, Japan}
\affiliation{Tokyo College and Department of Applied Physics, University of Tokyo, Tokyo 113-8656, Japan}
\author{J. D. Reim}%
\affiliation{Institute of Multidisciplinary Research for Advanced Materials (IMRAM), Tohoku University, Katahira 2-1-1, Sendai 980-8577, Japan.}
\author{Y. Nambu}%
\affiliation{Institute for Materials Research (IMR), Tohoku University, Katahira 2-1-1, Sendai 980-8577, Japan.}
\affiliation{Organization for Advanced Studies, Tohoku University, Sendai 980-8577, Japan}
\affiliation{FOREST, Japan Science and Technology Agency, Kawaguchi, Saitama 332-0012, Japan}
\author{T. J. Sato}%
\affiliation{Institute of Multidisciplinary Research for Advanced Materials (IMRAM), Tohoku University, Katahira 2-1-1, Sendai 980-8577, Japan.}

\date{\today}% It is always \today, today,

\begin{abstract}
Lattice formation of swirling textures is ubiquitous in solid-state materials, such as a magnetic skyrmion lattice in chiral magnets.  
In the magnetic skyrmion lattices, their moving states and dynamics under external perturbations are still unrevealed, 
although a detailed understanding of the dynamics is crucial to realizing spintronic applications, such as magnetic domain-wall racetrack memory~\cite{Parkin2008,Fert2013}.  
Here, we report in detail on the transient state of a moving magnetic skyrmion lattice in bulk single-crystalline MnSi under alternating current (AC) 
using small-angle neutron scattering.  
A rotation and concomitant broadening of the spot width in the azimuthal direction of the magnetic skyrmion reflections 
originating from the plastic deformation of the magnetic skyrmion lattice were found only at low AC frequencies~\cite{Okuyama2019}, 
whereas above the threshold AC frequency ($f_{\mathrm{t}}$ $\sim$ 0.12~Hz) the rotation was not observed, and the spot width becomes sharper.  
The observed complex response of the magnetic skyrmion reflections can be explained 
by the change in dislocation density in the magnetic skyrmion lattice.  
At frequencies higher than $f_{\mathrm{t}}$, the magnetic skyrmions oscillate removing the dislocations, indicating that the dislocation density is controlled by the AC frequency.  
\end{abstract}

%\pacs{75.70.-i, 64.75.St, 75.47.Lx}% PACS, the Physics and Astronomy
                              
\maketitle

\section{Introduction}

Swirling textures, called topological defects, characterized by discrete topological numbers, are ubiquitous 
in various scientific fields~\cite{Mermin1979,Blatter1994,Tokura2021,Wen1989}.  
The dynamics of moving topological defects by an external perturbation is one of the most fascinating scientific topics and has been investigated for 
decades~\cite{Fert2013,Anderson1966,Ruutu1997,Yaron1995,Braun1996,Gordeev1997,Jonietz2010,Zang2011,Schulz2012,Yu2012,Nagaosa2013,Jiang2017,Litzius2017}.  
The moving lattice states of topological defects in solid-state materials have been thoroughly investigated in quantum vortices in type-II 
superconductors~\cite{Yaron1995,Gurp1968,Matsuda1996,Duarte1996,Marchevsky1997,Pardo1998,Troyanovski1999,Kolton1999,Togawa2000}.  
Various moving lattice states, such as the plastic flow and coherently moving Bragg lattice of quantum vortices, 
have been discussed~\cite{Balents1998,Doussal1998,Olson1998}.  
In sharp contrast, to the best of our knowledge, few investigations to observe the transient state of the moving quantum vortices 
were conducted by microscopically probing the periodicity of the quantum vortices, such as by neutron scattering, 
because the scattering cross-section from the quantum vortices consisting of tiny magnetic flux 
is considerably small~\cite{Yaron1995,Matsuda1996}.  

Skyrmions, topological defects originally predicted in the nonlinear field theory, can now be found in a broad class of 
magnetic materials~\cite{Skyrme1961,Derrick1964,Bogdanov1989}.  
The magnetic skyrmions often condense into a triangular lattice, which was first discovered experimentally as a six-fold magnetic Bragg peak 
by small-angle neutron scattering (SANS) in the chiral magnet MnSi~\cite{Mublbauer2009}.  
Magnetic skyrmions have attracted significant attention because of 
the following prominent characteristics: 
(i) topological protection; the magnetic skyrmion can hardly be annihilated once it is created, and (ii) spin-transfer torque, 
spin dynamics of skyrmion is strongly coupled to spin current or electric current flow in metallic materials.  
Owing to these characteristics, the potential for controlling magnetic skyrmions 
and ultimately their potential application in spintronics has been concluded~\cite{Fert2013}.  
Thus, magnetic skyrmions have been extensively investigated to elucidate their dynamics under external perturbations~\cite{Fert2013,Okuyama2019,Jonietz2010,Zang2011,Schulz2012,Yu2012,Everschor2012,Lin2013,Iwasaki2013,Iwasaki2014,Reichhardt2015,Zhang2018,Yokouchi2018,TakuroSato2019}.  
As a pioneering work on revealing the motion of the magnetic skyrmion lattice by an electric current flow, 
a SANS experiment with an intentionally applied temperature gradient was conducted and clarified 
that the magnetic skyrmion lattice homogeneously rotates in the entire sample area 
by the spatially inhomogeneous spin-transfer torque~\cite{Jonietz2010,Everschor2012}.  
The skyrmion-lattice motion caused only by the electric current flow was also observed in the Hall resistivity measurement 
under thermally homogeneous conditions~\cite{Schulz2012}, 
and then the SANS measurement under the homogeneous conditions displayed the plastic deformation behavior of moving magnetic skyrmion lattices~\cite{Okuyama2019}.  
These experimental findings indicate that the electric current flow induces skyrmion-lattice motion in MnSi 
above the threshold current density $j_{\mathrm{t}} \sim$ 1~MA/m$^{2}$ (10$^6$~A/m$^{2}$), which is significantly smaller than the one required 
for the motion of magnetic domain boundaries, where the threshold value is approximately 
$j \leq$ 1~GA/m$^{2}$ (10$^9$~A/m$^{2}$)\cite{Myers1999,Grollier2003,Tsoi2003,Yamanouchi2004,Baltz2018}.  
Furthermore, the scattering cross-section from the magnetic skyrmion lattice in MnSi is larger than that from the quantum vortex lattice 
in type-II superconductors, simply because of the large magnetic moments of manganese in MnSi.  
Thus, the magnetic skyrmion is a promising candidate for investigating the transient lattice deformation of moving topological defects in solids.  
Additionally, the movement of the lattice in a spatially inhomogeneous and temporally changing manner is of further interest, 
and experimental proof for such a circumstance is desired.  

Here, we investigated the transient deformation process of a moving magnetic skyrmion lattice in MnSi in response to an alternating current (AC) 
using SANS under thermally homogeneous conditions.  
A rotation and concomitant broadening of the spot width in the azimuthal angle direction of the magnetic skyrmion reflections were observed 
above $j_{\mathrm{t}}$ and below the threshold AC frequency ($f_{\mathrm{t}}$), 
in accord with previously reported experiments under the direct current (DC)~\cite{Okuyama2019}.  
In stark contrast, above $f_{\mathrm{t}}$, the spot rotation is suppressed, and the spot width becomes even sharper compared to the pristine state.  
This article describes and discusses the transient process of the lattice deformation of the magnetic skyrmion and its slow dynamics.  

\section{Experimental}

\begin{figure}
\begin{center}
\includegraphics*[width=80mm,clip]{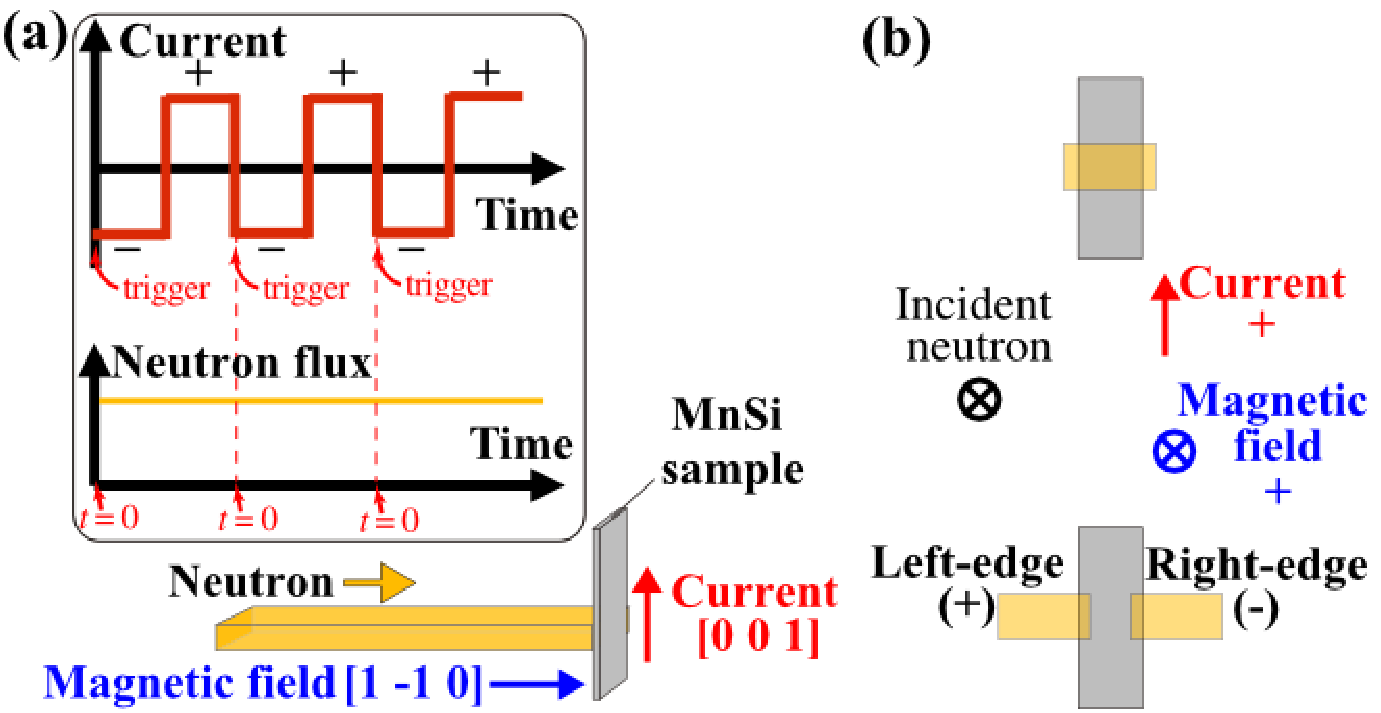}
\end{center}
\caption{\label{fig_01} 
Experimental setting of small-angle neutron scattering (SANS) for an alternating current (AC).  
(a) Schematic drawing of the MnSi sample and experimental setting for AC flow.  
To eliminate the temporal temperature variation inside the sample, we selected the AC with a square waveform.   
(b) Schematic illustration of the incident-neutron-illumination area for investigating the peak broadening 
at the entire horizontal cross-section of the sample (top) and the spatially inhomogeneous rotation at the sample edges (bottom) of the skyrmion lattice reflections.  
}
\end{figure}

The MnSi single crystal was grown along the [1 0 0] direction using the Czochralski method.  
The sample quality was checked by electric resistivity measurements.  
In our MnSi sample, the residual resistivity ratio was $\sim$50 and $T_{\mathrm{C}}$ was $\sim$29~K.  
The single crystal samples were cut in a rectangular shape of 1.4~mm (width) $\times$ 7.5~mm (height) $\times$ 0.4~mm (thickness) 
for the SANS experiment.  
The SANS experiments were performed at NG7 (National Institute of Standards and Technology).  
The incident neutron wavelength was $\lambda_{\rm i}$ = 6~\AA\ with $\rm \Delta \lambda / \lambda_{\rm i}$ = 14 \%.  
The experimental setting was identical to that used in our previous study~\cite{Okuyama2019}.  
To further suppress the temperature inhomogeneity in the measured region and check the sample position dependence, 
only a tiny part of the sample was illuminated by a narrow neutron beam of the cross-section being 2.0~mm (width) $\times$ 1.0~mm (height).  
The illumination area of the incident neutron for investigating the peak broadening behavior (top) and the spatially inhomogeneous rotation 
at the sample edges (bottom) are shown in Fig.~\ref{fig_01}(b).  
The MnSi sample was mounted on a Cu sample cell and installed in a horizontal-field magnet with a magnetic field applied 
along [1 -1 0] parallel to the incident neutron.  
To observe the magnetic skyrmion lattice, we applied a magnetic field $B_{\mathrm{ext}}$ = 0.2~T, whose value is 
the external magnetic field without correcting for the magnetic permeability of the MnSi sample.  
An electric current for DC and AC with a square waveform of up to 2.0~A ($j$ = 3.6~MA/m$^{2}$) was applied along the [0 0 1] direction.  
The temperature gradient in the MnSi sample, along and perpendicular to the current flow direction was confirmed 
to be less than 0.035~K/mm at $j$ = 2.7~MA/m$^{2}$~\cite{Okuyama2019}.  

\section{Experimental results and analyses}

\begin{figure}
\begin{center}
\includegraphics*[width=70mm,clip]{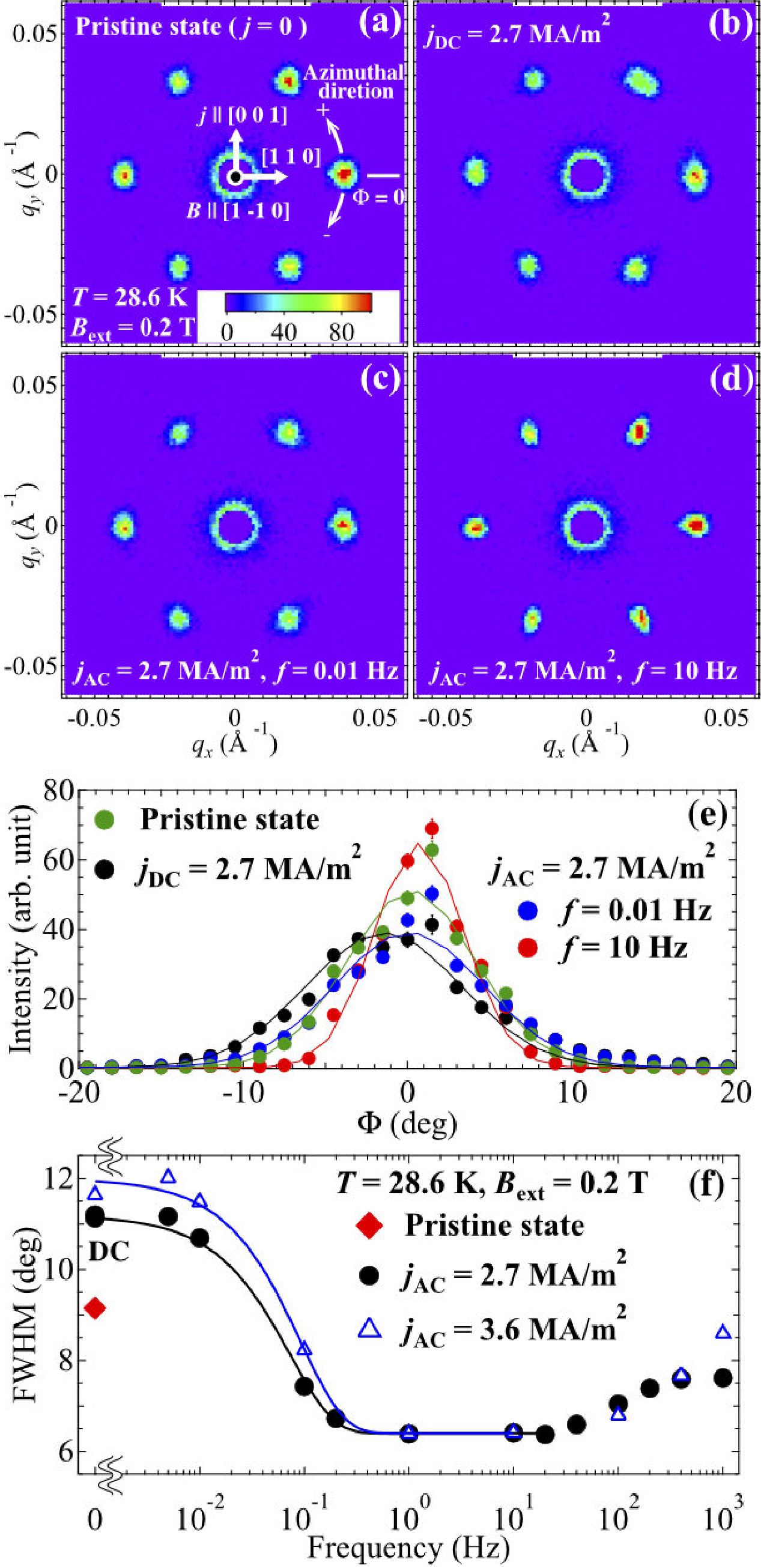}
\end{center}
\caption{\label{fig_02} 
SANS data for investigating the peak broadening of the skyrmion reflection under AC flow.  
SANS measured at $T$ = 28.6~K under $B_{\mathrm{ext}}$ = 0.2~T.  
In these measurements, the setting at the top of Fig.~\ref{fig_01}(b) was used.  
The skyrmion reflections were measured at the pristine state ($j$ = 0) (a) and under the direct current (DC) 
at $j_{\mathrm{DC}}$ = 2.7~MA/m$^{2}$ (b), and the AC at $j_{\mathrm{AC}}$ = 2.7~MA/m$^{2}$ 
for the AC frequency $f$ = 0.01 (c) and 10~Hz (d), respectively.  
The azimuthal angle direction ($\Phi$) is indicated by the arrow in panel (a).  
Here, the counterclockwise rotational direction is positive and the position of the white bar is defined as zero position.  
All data were measured for 10~minutes.  
(e) The azimuthal angle dependences of the integrated intensity near $\Phi$ = 0 obtained from SANS data.  
The lines are fitting results of the Gaussian functions.  
Hereafter, we use 1$\sigma$ standard deviation as error bar.  
(f) Frequency dependences of the azimuthal spot widths for the skyrmion reflections at $j_{\mathrm{AC}}$ = 2.7~MA/m$^{2}$ (closed circle) 
and $j_{\mathrm{AC}}$ = 3.6~MA/m$^{2}$ (open triangle).  
Red diamond stands for the azimuthal spot width for the pristine state.  
The black and blue lines are guides to the eye for the data at $j_{\mathrm{AC}}$ = 2.7 and 3.6~MA/m$^{2}$, respectively.  
}
\end{figure}

Representative time-integrated SANS patterns measuring the broadening of the spot width of the magnetic skyrmion reflection at $T$ = 28.6~K 
and $B_{\mathrm{ext}}$ = 0.2~T are shown in Figs.~\ref{fig_02}(a)-\ref{fig_02}(d).  
In this setting, the incident neutron illuminates the entire horizontal cross-section of the sample [see Fig.~\ref{fig_01}(b) (top)].  
Hereafter, the densities of the DC and AC are represented as $j_{\mathrm{DC}}$ and $j_{\mathrm{AC}}$, respectively.  
Figures~\ref{fig_02}(a)~and~\ref{fig_02}(b) display the six-fold reflections from the magnetic skyrmion lattice in the pristine state ($j$ = 0) 
and $j_{\mathrm{DC}}$ = 2.7~MA/m$^{2}$, respectively.  
In the setting of the DC flow, broadening of the azimuthal spot width above $j_{\mathrm{t}}$ was observed, 
which is consistent with the results of our previous work~\cite{Okuyama2019}.  
In Figs.~\ref{fig_02}(c)~and~\ref{fig_02}(d), the SANS data at $j_{\mathrm{AC}}$ = 2.7~MA/m$^{2}$ with the AC frequencies of 0.01 and 10~Hz 
are respectively shown.  
The representative azimuthal angle dependence of the integrated intensity of the magnetic skyrmion reflection near $\Phi$ = 0 
obtained from the SANS data is shown in Fig.~\ref{fig_02}(e).  
In comparison to the pristine state, the azimuthal spot width is sharper at 10~Hz.  

To compare quantitatively the spot widths, the AC frequency dependences of the azimuthal spot width obtained by the six Gaussian function fitting 
to the six-fold magnetic skyrmion reflections at $j_{\mathrm{AC}}$ = 2.7 and 3.6~MA/m$^{2}$ are shown in Fig.~\ref{fig_02}(f) 
[The details for the data analyses are shown in Fig.~\ref{fig_06}, Appendix~A].  
The data can be divided into two frequency regions.  
For frequencies lower than 0.1~Hz, the azimuthal spot width is broader than in the pristine state.  
This peak broadening is consistent with the results obtained under DC~\cite{Okuyama2019}.  
In contrast, for $f \geq$ 1~Hz, the magnetic skyrmion reflections do not exhibit peak broadening, and the azimuthal spot width is sharper than in the pristine state.  
The spot width remained sharp up to 100~Hz and slightly increased thereafter.  

\begin{figure}
\begin{center}
\includegraphics*[width=78mm,clip]{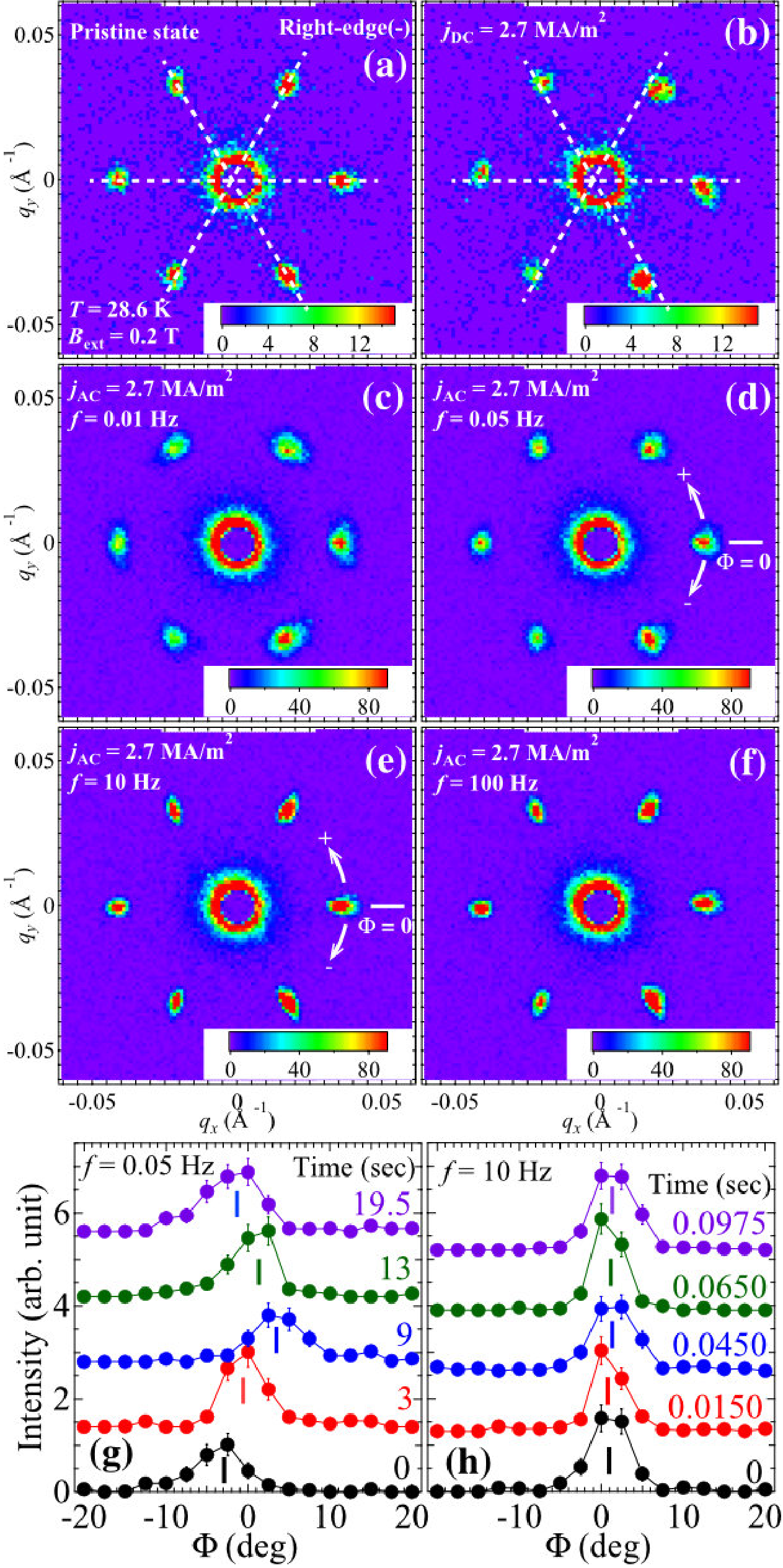}
\end{center}
\caption{\label{fig_03} 
SANS data for investigating the rotation of the skyrmion reflections at the sample edge under AC flow.  
The SANS patterns measured at the right edge (-) of the sample as schematically illustrated in Fig.~\ref{fig_01}(b) (bottom) are shown.  
The measurements were conducted at the pristine state (a), $j_{\mathrm{DC}}$ = 2.7~MA/m$^{2}$ (b), 
and $j_{\mathrm{AC}}$ = 2.7~MA/m$^{2}$ for $f$ = 0.01 (c), 0.05 (d), 10 (e), and 100~Hz (f), respectively.  
All measurements were conducted at $T$ = 28.6~K and $B_{\mathrm{ext}}$ = 0.2~T.  
The data for (a) and (b) were measured for 20~minutes.  
The white dotted lines in (a) and (b) are guides to the eye for the peak positions of the skyrmion reflections in pristine condition.  
For the data (c-f), a 60-min measurement was conducted and the AC flow was applied.  
(g,~h) The time dependences of the integrated intensity near $\Phi$ = 0 obtained from SANS data for the skyrmion reflections 
at the AC frequency $f$ = 0.05 (g) and 10~Hz (h), respectively.    
The bars in (g) and (h) stand for the peak center positions obtained by Gaussian function fitting for each time.  
Each data set is shifted vertically to improve readability.  
}
\end{figure}

To determine the origin of the peak broadening of the magnetic skyrmion reflection, we also measured the SANS patterns 
at the left and right edges of the sample to investigate the spatially inhomogeneous rotation of the magnetic skyrmion reflections in the azimuthal angle direction, 
as shown in Fig.~\ref{fig_01}(b) (bottom).  
In Figs.~\ref{fig_03}(a)-\ref{fig_03}(f), representative time-integrated SANS patterns measured only at the right edge are shown 
[The SANS data and analyses at the left edge are shown in Figs.~\ref{fig_08} and~\ref{fig_10}, Appendix~C].  
Figures~\ref{fig_03}(a)~and~\ref{fig_03}(b) show the magnetic skyrmion patterns in the pristine state and $j_{\mathrm{DC}}$ = 2.7~MA/m$^{2}$, respectively.  
The magnetic skyrmion reflections exhibited clockwise rotations at $j_{\mathrm{DC}}$ = 2.7~MA/m$^{2}$, 
originating from the plastic deformation caused by the electric current flow, as discussed in the previous work~\cite{Okuyama2019}.  
As shown in Figs.~\ref{fig_03}(c)~and~\ref{fig_03}(d), the broadening of the spot widths was observed at $j_{\mathrm{AC}}$ = 2.7~MA/m$^{2}$ with $f$ = 0.01 and 0.05~Hz.  
At these frequencies, MnSi exhibited rotation in the azimuthal angle direction of the magnetic skyrmion reflections near the sample edges.  
Thus, this peak broadening originates from the superimposition of clockwise and counterclockwise rotated reflections induced by the temporally changing AC flow.  
In contrast, the azimuthal spot widths for $f$ = 10 and 100~Hz in Figs.~\ref{fig_03}(e)~and~\ref{fig_03}(f) did not display peak broadening, 
which is consistent with the results in the setting for the whole-sample illumination, as shown in Fig.~\ref{fig_02}(d).  

Next, we show the time evolution of the SANS data to investigate the transient state of rotation and peak broadening 
of the magnetic skyrmion reflections.  
Here, we used the time-resolved SANS data.  
The time-resolved SANS intensity was integrated by repeating the inversion of the electric current.
We defined `Time = 0' at the falling edge of the electric current with a square waveform, as shown in Fig.~\ref{fig_01}(a).  
In Fig.~\ref{fig_03}(g), the time evolution of the azimuthal angle dependency of the integrated intensity of the magnetic skyrmion reflection 
near $\Phi$ = 0 at $f$ = 0.05~Hz is shown.  
The center position of the magnetic skyrmion reflection at 0~s was approximately -3$^{\circ}$.  
After 3~s, the magnetic skyrmion reflection moved positively, and the spot width appeared sharper than that at 0~s.  
After 9~s, the center position remained positive approximately at 3$^{\circ}$ and the spot width became broader than that at 3~s.  
At 10~s, the electric current was switched from a negative to a positive flow direction. 
Then, the magnetic skyrmion reflection moved in the negative direction at 13~s, which is 3~s after the polarity change, 
and the spot width became sharper again.  
The center position at 19.5~s later returned to the original approximately at -3$^{\circ}$, and the spot width became broader again.  
In stark contrast, the center position of the magnetic skyrmion reflection measured at $f$ = 10~Hz did not display a drastic change, 
and the spot width also remained sharp [see Fig.~\ref{fig_03}(h)].  

\begin{figure*}
\begin{center}
\includegraphics*[width=160mm,clip]{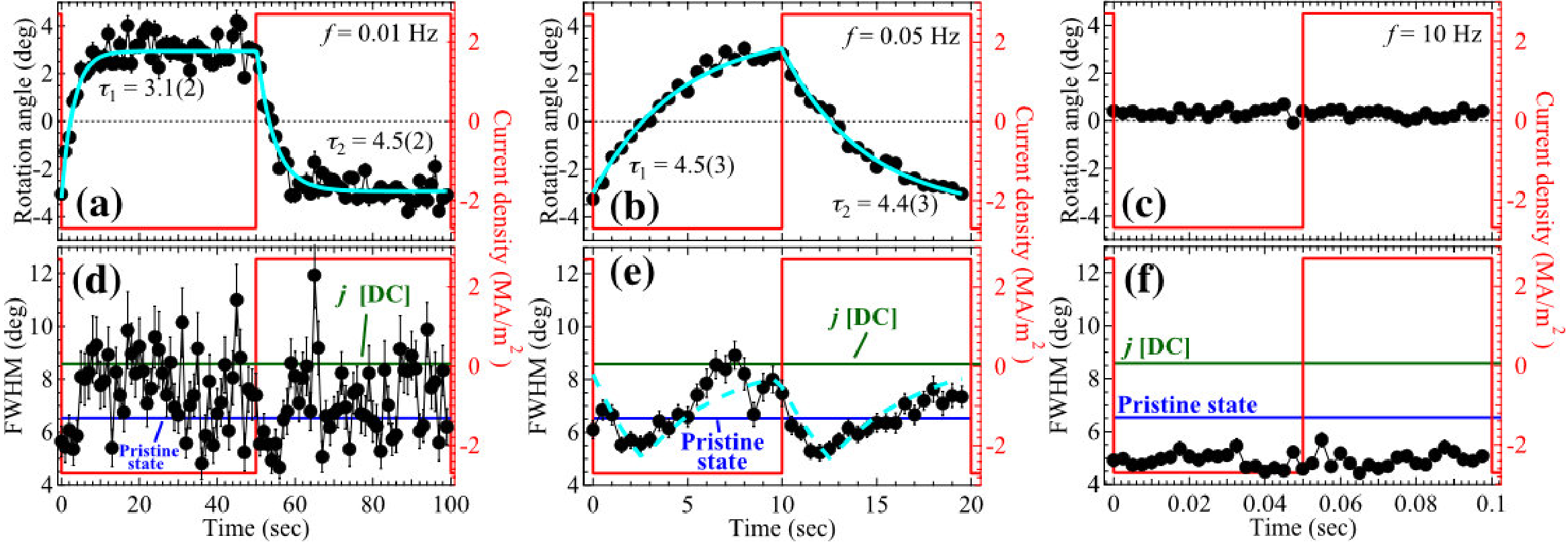}
\end{center}
\caption{\label{fig_04} 
Frequency-dependent time evolution of the rotation angle of the center position and azimuthal spot width of the skyrmion reflections 
measured at the right edge of the sample under AC flow.  
The time evolution of the rotation angle and the spot widths of the magnetic skyrmion reflections 
obtained from the six Gaussian function fitting to the six-fold magnetic skyrmion reflections measured at $j_{\mathrm{AC}}$ = 2.7~MA/m$^{2}$ 
for $f$ = 0.01 (a,~d), 0.05 (b,~e), and 10~Hz (c,~f), respectively.  
Red lines stand for the time dependence of the applied electric-current density with the square waveform.   
Cyan lines in panels (a) and (b) represent the fitting results of the Debye relaxation equation.  
The dotted cyan line in panel (e) is the absolute value of the rotation angle of the magnetic skyrmion reflection in panel (b) multiplied 
by the constant value and added the background constant.  
Dark blue and green lines in panel (d-f) stand for the azimuthal spot width for the pristine state 
and $j_{\mathrm{DC}}$ = 2.7~MA/m$^{2}$, respectively.  
}
\end{figure*}

To quantitatively analyze the time evolution of the magnetic skyrmion reflections under AC flow, 
the azimuthal angle-dependent integrated intensity of the magnetic skyrmion reflections obtained from the time-resolved SANS data 
was fitted to six equally distributed Gaussian functions.  
The obtained time dependence of the rotation angle of the center position and azimuthal spot width 
of the magnetic skyrmion reflections measured at the right edge of the sample, are shown in Figs.~\ref{fig_04}(a)-\ref{fig_04}(f).  
The rotation angle for $f$ = 0.01 and 0.05~Hz in Figs.~\ref{fig_04}(a)~and~\ref{fig_04}(b) displays the drastic time evolution.  
After inverting the electric current from negative (positive) to positive (negative), the center position of the rotation angle direction 
changed from positive (negative) to negative (positive) with a finite relaxation time.  
To evaluate the relaxation time, the time dependence of the center position of the rotation angle was fitted 
to the following Debye relaxation equation 
[cyan lines in Figs.~\ref{fig_04}(a)~and~\ref{fig_04}(b)]: 
\begin{equation}
\label{eq_01}
\Phi(t) = \left\{
\begin{array}{ll}
A - B \times \exp{( -t / \tau_{1} )} & (0 \leqq t < 1/2f)\\
- A + B \times \exp{( -(t - 1/2f) / \tau_{2} )} & (1/2f \leqq t < 1/f), 
\end{array}
\right.
\end{equation}  
where $\Phi(t)$, $f$, and $t$ represent the rotation angle of the magnetic skyrmion reflection, AC frequency, and time, respectively.  
$A$ and $B$ are constant values, and $\tau_{1}$, and $\tau_{2}$ are the relaxation times.  
The obtained relaxation times were as follows: 
$\tau_{1}$ = 3.1(2)~s and $\tau_{2}$ = 4.5(2)~s for $f$ = 0.01~Hz, and $\tau_{1}$ = 4.5(3)~s and $\tau_{2}$ = 4.4(3)~s for $f$ = 0.05~Hz.  
The average relaxation time was $\tau_{av} \sim$ 4.1(1)~s.  
Hereafter, we defined the threshold frequency $f_{\mathrm{t}}$ as $1/(2 \tau_{av}) \sim 0.12$~Hz.  
In contrast, for the data at $f$ = 10~Hz, as shown in Fig.~\ref{fig_04}(c), where the periodicity of the electric-current inversion 
was shorter than the relaxation time of the rotation of the magnetic skyrmion reflection, the temporal change of the rotation angle 
was not observed at all.  

Furthermore, as shown in Figs.~\ref{fig_04}(d)-\ref{fig_04}(f), the azimuthal spot widths of the magnetic skyrmion reflections display 
a more complicated time evolution.  
For $f$ = 0.05~Hz, 1.5~s later after inverting the electric current flow direction, the azimuthal spot width remained at $\sim$5$^{\circ}$ for the next 2~s.  
After 4~s, the azimuthal spot width gradually increased with time.  
Then, the azimuthal spot width saturated near the value observed under DC flow until the electric current direction was reinverted.  
For $f$ = 0.01~Hz, the intriguing time evolution of the azimuthal spot width follows the result for $f$ = 0.05~Hz; 
however, the statistic was not sufficient to discuss the small change of the azimuthal spot width by the time evolution.  
For $f$ = 10~Hz, the azimuthal spot width remained less than 
5$^{\circ}$ without an observable time evolution, 
which was sharper than that for the pristine states.  

\section{Discussion}

In this section, we discuss a plausible model for understanding the time evolution of the SANS pattern under AC flow.  
In the previous study on DC flow~\cite{Okuyama2019}, we used the plastic flow model to explain 
the spatially inhomogeneous rotation of magnetic skyrmion reflection, where magnetic skyrmions move with dislocations 
of the magnetic skyrmion lattices slipping along the domain boundaries.  
We should consider whether this model would apply to the SANS results under AC flow.  
In the frequency dependence of the azimuthal spot width in Fig.~\ref{fig_02}(f), 
the broadening of the azimuthal spot width originating from plastic deformation was observed only below $f_{\mathrm{t}}$.  
Thus, we should separately discuss the origins of the observed phenomena below and above $f_{\mathrm{t}}$.  

\begin{figure*}
\begin{center}
\includegraphics*[width=130mm,clip]{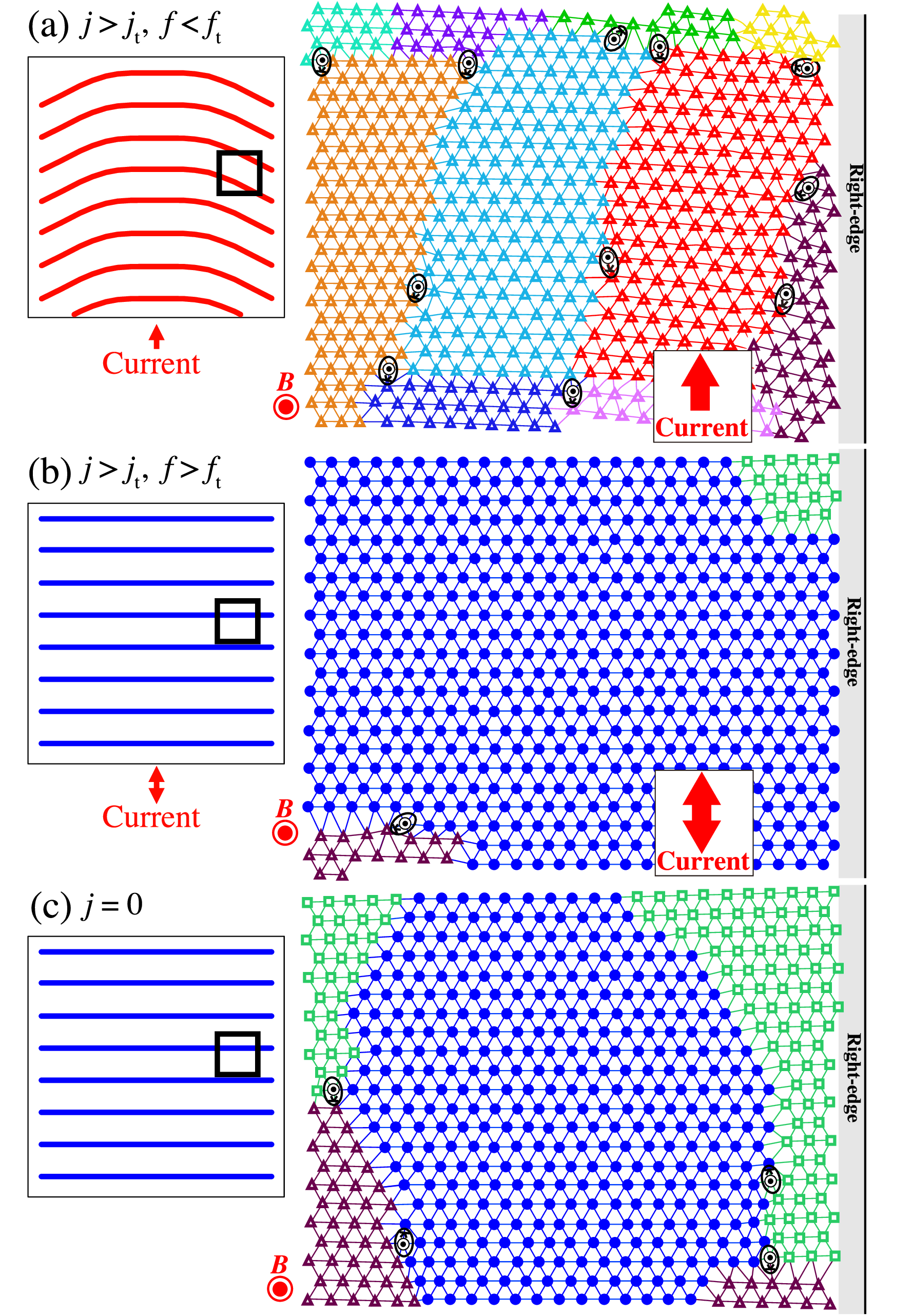}
\end{center}
\caption{\label{fig_05} 
Schematics for top view of the bends, domain structure, and the dislocations of the magnetic skyrmion lattice near the sample edge.  
The left figures schematically represent the bends of the magnetic skyrmion lattice to the electric current flow in the whole sample 
above the threshold electric current density at $f < f_{\mathrm{t}}$ (a) and $f > f_{\mathrm{t}}$ (b), and in the pristine state (c).  
The right figures schematically represent the magnified areas enclosed by the black squares in the left figures.  
The domain structure and the dislocations of the magnetic skyrmion lattice near the right edge of the sample are depicted.  
The different domains are represented by different colors.  
The solid and open symbols stand for the magnetic skyrmions.  
The open triangles (squares) mean the clockwise (counterclockwise) rotated magnetic skyrmion domain, respectively.  
The blue solid circles stand for the unrotated ($\phi$ = 0) magnetic skyrmion domain.  
The black double circles stand for the end of the extra lattice plane on the dislocation of the magnetic skyrmion lattice.  
The $\veebar$ and $\barwedge$ symbols stand for the directions of the fault lines and Burgers vector.  
The dislocation in (c) is thermally induced by the first-order phase transition, and thus the direction of the fault lines of the dislocation is random.  
}
\end{figure*}

Here, we extend the plastic flow model to explain the complex time evolution of the SANS data below $f_{\mathrm{t}}$.  
First, we consider that the magnetic skyrmion lattices are sheared by the plastic deformation at the sample edge above $j_{\mathrm{t}}$.  
The bends of the magnetic skyrmion lattices to electric current flow gradually relax from the sample edge to the sample center, 
as shown schematically in Fig.~\ref{fig_05}(a).  
In that case, the domain of the magnetic skyrmion lattice along the perpendicular direction of the bending angle should be fragmented into smaller domains 
with increasing the bending angle.  
The dislocation density in the magnetic skyrmion lattice is at maximum when the domain size of the magnetic skyrmion lattice is minimal.  
It is expected that the correlation length in the azimuthal angle direction (the rotational direction) is inversely proportional 
to a derivative $d|\phi(r)|$/$dr$, where $\phi(r)$ is the bending angle at the distance $r$ from the sample edge.  
The bending behavior of the condensate near the sample edge under an electric current flow has also been discussed 
in charge-density-wave (CDW) systems~\cite{comment01}.  

Next, to validate the correctness of this model, we investigate the relation of the experimentally obtained correlation length 
in an azimuthal angle direction and $d|\phi(r)|$/$dr$ of the magnetic skyrmion.  
In the SANS experiment, the broadening of the spot width and the rotation in the azimuthal angle direction of the magnetic skyrmion reflections are separately observed.  
Note that the azimuthal spot width and the rotation of the magnetic skyrmion reflection are respectively the reciprocal of the correlation length 
in the azimuthal angle direction and $d|\phi(r)|$/$dr$ averaged in the neutron illumination area.  
Therefore, if the above model is correct, the spot width in the azimuthal angle direction should be proportional to 
the absolute value of the rotation angle $| \Phi |$ of the magnetic skyrmion reflections as follows: 
Spot width  = $a_0$ + $b$ $| \Phi(t) |$, where $a_0$ and $b$ are constant values.  
Time-dependence of the spot width estimated assuming this equation is shown by a dotted cyan line in Fig.~\ref{fig_04}(e), 
which is to be compared with the experimentally observed time dependence.  
A good agreement to the experimental result can be seen in the figure.  
Notably, the time evolution of the rotation angle of the magnetic skyrmion reflection was well-fitted by the Debye relaxation model 
with a slow relaxation time, which is anticipated to originate from the motion of a large assembly.  
The moving magnetic skyrmions accompanying the rotation of the large domains and the increase or decrease in the dislocation density 
in the magnetic skyrmion lattice in the extended plastic-flow model are compatible with the motion of such a large assembly.  
From this discussion, we conclude that the extended plastic-flow model can explain the observed phenomena below $f_{\mathrm{t}}$.  

Above $f_{\mathrm{t}}$ under AC flow, as the rotation of the magnetic skyrmion reflection is suppressed, 
we consider that the magnetic skyrmions do not move long distances but oscillate within short displacements.  
Therefore, we recall the random organization discussed in a periodically driven system, 
where the particles are self-organized to avoid collisions after repeatedly colliding with neighboring particles~\cite{Corte2008}.  
The magnetic skyrmions under AC flow above $f_{\mathrm{t}}$ oscillate cyclically by pushing each other, similar to the situation 
in the earlier model simulation for the periodically driven system.  
Then, we speculate that the magnetic skyrmions are rearranged equidistant and the dislocations in the magnetic skyrmion lattice are removed by this effect.  
Consequently, the azimuthal spot width of the magnetic skyrmion reflection becomes sharp 
because the domain size of the magnetic skyrmion lattice is enlarged, as shown in Fig.~\ref{fig_05}(b).  
This type of random organization effect in the topological matter has also been discussed 
in the quantum vortex of type-II superconductors~\cite{Okuma2011,Okuma2012}.  
In stark contrast, however, the rearrangement of the magnetic skyrmions should not originate from the thermal vibration effect 
since the plastic deformation of the magnetic skyrmion lattice exhibits a long-time memory effect~\cite{Okuyama2019}.  
It is also noteworthy that the azimuthal spot width of the magnetic skyrmion reflection above $f_{\mathrm{t}}$ becomes even sharper than in the pristine state.  
To understand these intriguing characteristics, it may be crucial that the domain size of the magnetic skyrmion lattice in the pristine state is not large.  
Many dislocations would exist at the domain boundaries owing to the first-order phase-transition nature of the magnetic skyrmion phase in MnSi 
and disturb the extension of the domain size of the magnetic skyrmion lattice, as depicted in Fig.~\ref{fig_05}(c).  
Thus, it is reasonable that the azimuthal spot width is sharper than in the pristine state as the dislocations would be removed 
by the random organization effect for the cyclically oscillating magnetic skyrmion lattice.  

From the above discussion, we obtain vital information for spintronic applications, that is, the dislocation density 
in the magnetic skyrmion lattice in MnSi is controllable by AC frequencies.  
Whereas the dislocation density in the magnetic skyrmion lattice increases owing to the DC and AC flows below $f_{\mathrm{t}}$, 
the dislocation density decreases with the dislocations removed by the cyclic oscillation of the AC flow 
above $f_{\mathrm{t}}$, as clearly shown from our experimental findings.  
Furthermore, AC flow fabricates a cleaner magnetic skyrmion lattice than the thermally induced one.  
The manipulation method of the dislocation density in the magnetic skyrmion lattice discussed here will facilitate 
further research on the spintronic applications of topologically protected matter.  

Finally, it is noteworthy to estimate the velocity of the moving magnetic skyrmion lattice based on the above model.  
In MnSi, a decrease in the azimuthal spot width owing to the random organization effect was observed at $f$ $\sim$ 1~Hz near $f_{\mathrm{t}}$.  
It is suspected that the magnetic skyrmions oscillate with a short distance near $f_{\mathrm{t}}$, 
probably several skyrmions (less than 1,000~\AA\ for MnSi), for the practical emergence of the random organization effect.  
Because the magnetic skyrmions oscillate 1,000~\AA\ in 1~s, the velocity of the moving magnetic skyrmion lattice 
is approximately 10$^{-7}$~m/s.  
This velocity near $j_{\mathrm{t}}$ in MnSi is of the same order as the value in the noise measurement~\cite{TakuroSato2019}, 
but far from the value in the topological Hall measurement estimated as 10$^{-5}$~m/s~\cite{Schulz2012}.  
The difference in the velocity between our investigation and the result of the topological Hall measurement may originate from the difference 
in the observed magnetic skyrmions.  
The viscous magnetic skyrmion motion on the dislocation in the magnetic skyrmion lattice near the sample edge is highlighted in this SANS investigation, 
whereas in the topological Hall measurement the averaged magnetic skyrmion motion in the sample is measured.  

\section{Conclusion}

In summary, we measured the SANS patterns to investigate skyrmion lattice motion in bulk MnSi under AC flows.  
The azimuthal spot width of the magnetic skyrmion reflection displays peak broadening above $j_{\mathrm{t}}$ and below $f_{\mathrm{t}}$.  
Time-dependent variations for the spatially inhomogeneous rotation and azimuthal spot width of the magnetic skyrmion reflections 
were observed below $f_{\mathrm{t}}$ in stark contrast to the time-independent variation 
of the magnetic skyrmion reflections above $f_{\mathrm{t}}$.  
We explain these complex time-dependent responses of the magnetic skyrmion lattice to the AC flow with a bending of the magnetic skyrmion lattice 
and the change in its dislocation density, which is critical to understanding the dynamic transient states of the moving topological defect.  

\section*{Acknowledgements}

The authors thank J. S. White, H. M. R$\o$nnow, P. D. Butler, D. Higashi, R. Murasaki, K. Nawa, K. Yamauchi and T. Oguchi 
for the fruitful discussions.  
This work was in part supported by Grants-in-Aids for Scientific Research (No. 24224009, 26103006, 19K03709, 23K03311, 
18H03676, 19H01834, 19H05824, 21H03732, and 22H05145) 
from the Ministry of Education, Culture, Sports, Science and Technology (MEXT), Japan, 
by the Research Program for CORE lab of \textquotedblleft Dynamic Alliance for Open Innovation Bridging Human, 
Environment and Materials\textquotedblright\ in \textquotedblleft Network Joint Research Center for Materials and Devices\textquotedblright, 
and by the Japan Science and Technology Agency (Grant No. JPMJFR202V).  
Travel expense for the experiment was partly sponsored by the General User Program of ISSP-NSL, University of Tokyo.  

\section*{Appendix A: Data analysis methods for small-angle neutron scattering (SANS)}

\begin{figure}
\begin{center}
\includegraphics*[width=80mm,clip]{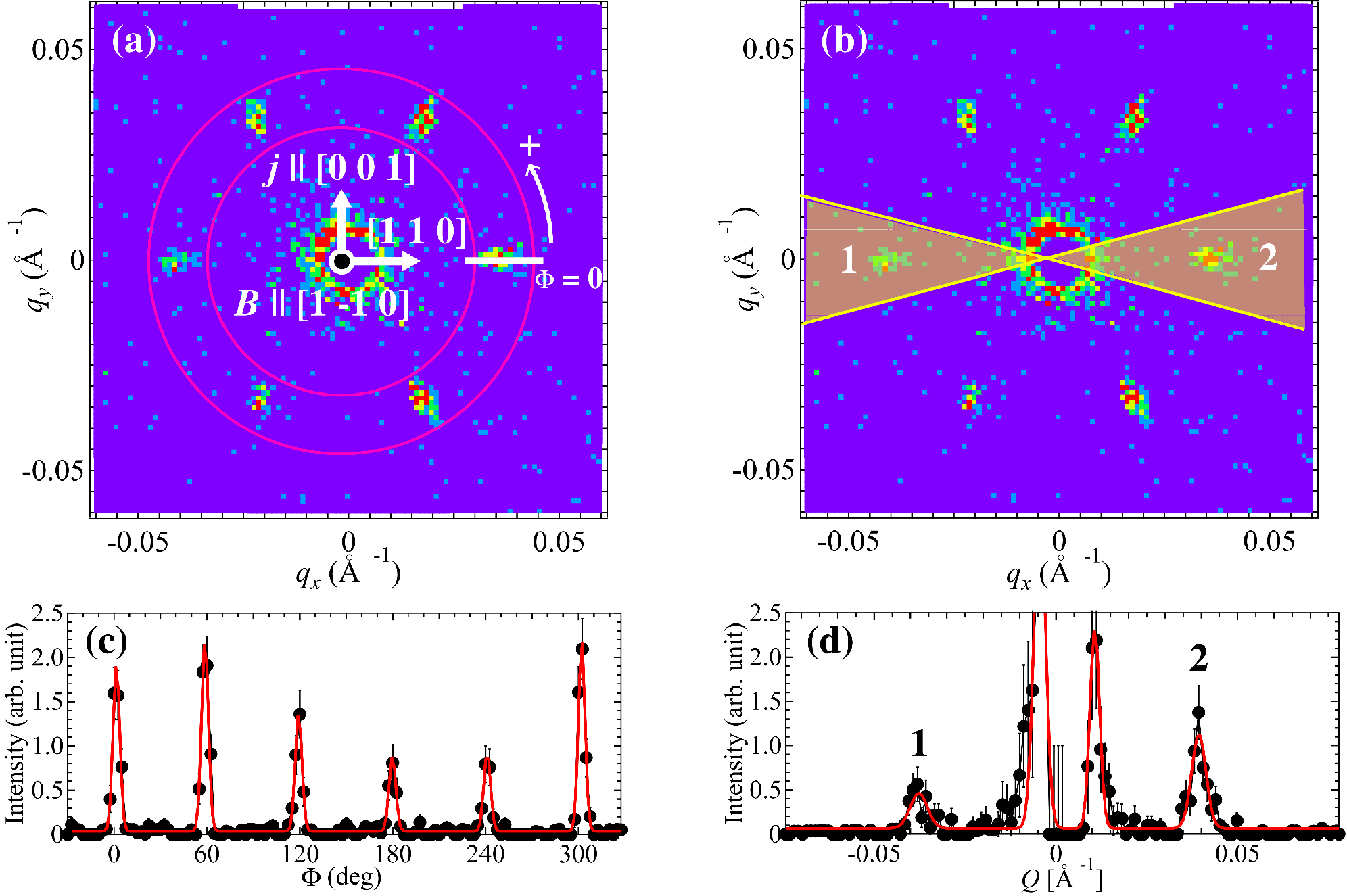}
\end{center}
\caption{\label{fig_06} 
Examples of the analyses of SANS data measured at $T$ = 28.6~K and $B_{\mathrm{ext}}$ = 0.2~T.  
(a) To obtain the azimuthal angle dependence of the SANS data, the intensity are integrated along the radial direction 
between the two red circles.  
The position indicated by the white line was defined as $\Phi$ = 0 and the counterclockwise direction was defined as positive.  
(b) To obtain the $\mathit{Q}$ dependence of the SANS data, the intensity are integrated along the azimuthal direction in the yellow colored area.  
The numbers of 1 and 2 indicate the corresponding skyrmion reflections in (d).  
(c) The azimuthal angle dependences of the intensity integrated along the radial direction between the two red circles.  
Red line shows the fitting result.  
From the fitting with six Gaussian functions, the peak position and spot width were obtained.  
(d) $\mathit{Q}$ dependence of the SANS intensities of the skyrmion reflections.  
From the fitting with four Gaussian functions, the peak position and spot width were obtained from the two peaks indicated by 1 and 2 numbers.  
}
\end{figure}

The analysis procedures are almost identical with our previous study~\cite{Okuyama2019}.  
To analyze the rotation angle and spot width broadening in the azimuthal angle direction of the magnetic skyrmion reflections 
as a function of frequency of an alternating current (AC), 
we obtained the azimuthal angle dependence of the scattering intensity from the SANS data by integrating along the radial direction 
between the two red circles shown in Fig.~\ref{fig_06}(a).  
The example of the azimuthal angle dependence of the integrated SANS intensity is shown in Fig.~\ref{fig_06}(c).  
All data were measured at $T$ = 28.6~K at $B_{\mathrm{ext}}$ = 0.2~T.  
Black filled circle and red line respectively stand for the obtained integrated intensity and the fitting curvature.  
The azimuthal angle dependence was fitted to six Gaussian functions and background intensity, 
from which the center position of the averaged rotation angle and full width at half maximum (FWHM) of the spot width 
of the magnetic skyrmion reflections were obtained.  
The detailed fitting function is defined as 
\begin{equation}
\label{eq_01}
I(\Phi) = \sum^{5}_{i = 0} A_i \times \exp{(-(\Phi-\Phi_0+60^{\circ}\times i)^2/2\sigma^2}) + \mathrm{BG}, 
\end{equation}  
where $I(\Phi)$, $\Phi$, $A_i$, and BG are the intensity, azimuthal angle, constant value, and background constant, respectively.  
$\Phi_0$ and $\sigma$ are the rotation angle and spot width of the skyrmion reflections, respectively.  

To analyze the variation of the spacing of the skyrmion lattice, the reciprocal lattice position $Q$ and spot width to $Q \hat{e}_{Q}$ 
direction of the skyrmion reflection were estimated.  
We obtained the $Q$ dependence of the intensity integrated along the azimuthal angle direction in the yellow colored area 
in Fig.~\ref{fig_06}(b), 
and the example of the $Q$ dependence of the intensities of the skyrmion reflections is shown in Fig.~\ref{fig_06}(d).  
The SANS intensity was fitted by four Gaussian functions, from which $Q$ position and spot width to $Q \hat{e}_{Q}$ direction were obtained.  
The fitting result is shown by the solid red line.  

\section*{Appendix B: Sequence to measure frequency dependences of SANS under AC flow}

\begin{figure*}
\begin{center}
\includegraphics*[width=160mm,clip]{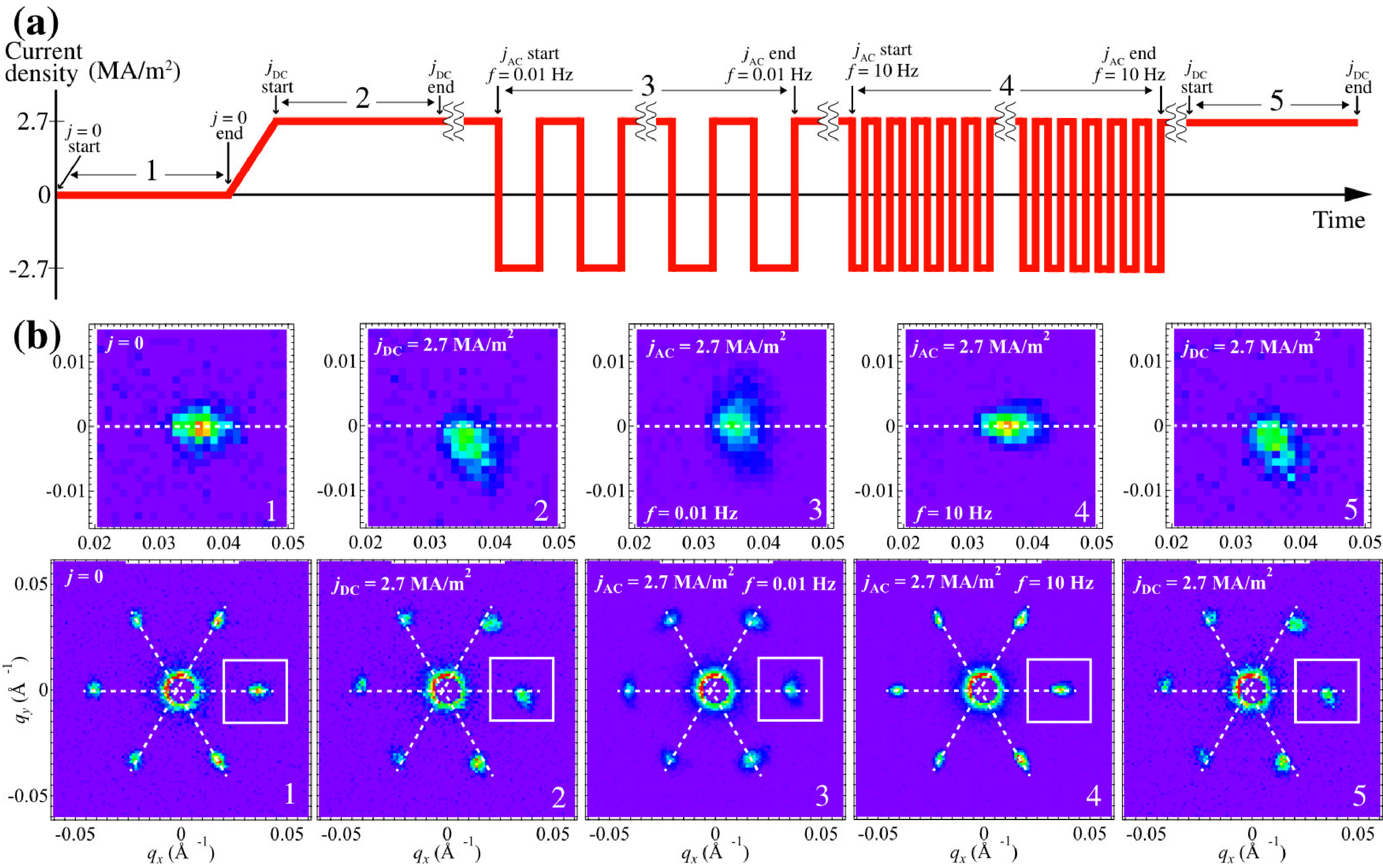}
\end{center}
\caption{\label{fig_07} 
(a) Schematics to explain the measurement procedure of the SANS data under the DC and AC.  
First, the SANS data for $j$ = 0 (1) was measured.  
Next, the electric current was gradually increased and the SANS data under the DC at $j_{\mathrm{DC}}$ = 2.7~MA/m$^{2}$ (2) was measured.  
Then, the SANS measurements under the AC (3 and 4) were conducted.  
The applied frequency was gradually increased from $f$ = 0.005~Hz to 1000~Hz.  
After measuring the SANS under AC, the SANS under DC (5) was measured again.  
(b) The examples for the SANS data at $T$ = 28.6~K and $B_{\mathrm{ext}}$ = 0.2~T measured at right edge of the sample are shown.  
The numbers indicate the corresponding SANS data measured in the region indicated by the numbers at (a).  
White dashed lines are guides to eye for the peak positions of the skyrmion reflections in pristine condition.  
Bottom and top figures show the whole magnetic skyrmion reflections and the single magnetic skyrmion reflection 
with magnifying the white solid line area in bottom figures, respectively.  
}
\end{figure*}

Here, we explain how to measure the AC frequency dependence of the SANS intensity.  
In Fig.~\ref{fig_07}, the schematic to explain the measurement procedure of the SANS data under AC are shown.  
At first, we measured the SANS data at the pristine state ($j$ = 0).  
Then, the electric current was gradually increased up to $j$ = 2.7 or 3.6~MA/m$^2$, and the measurement under the direct current (DC) was conducted.  
Next, the SANS measurements under the AC flow were conducted in order from lower to higher AC frequency.  
Finally, the SANS data was measured under DC flow again to confirm the reproducibility.  

\section*{Appendix C: Supplemental results}

\begin{figure}
\begin{center}
\includegraphics*[width=80mm,clip]{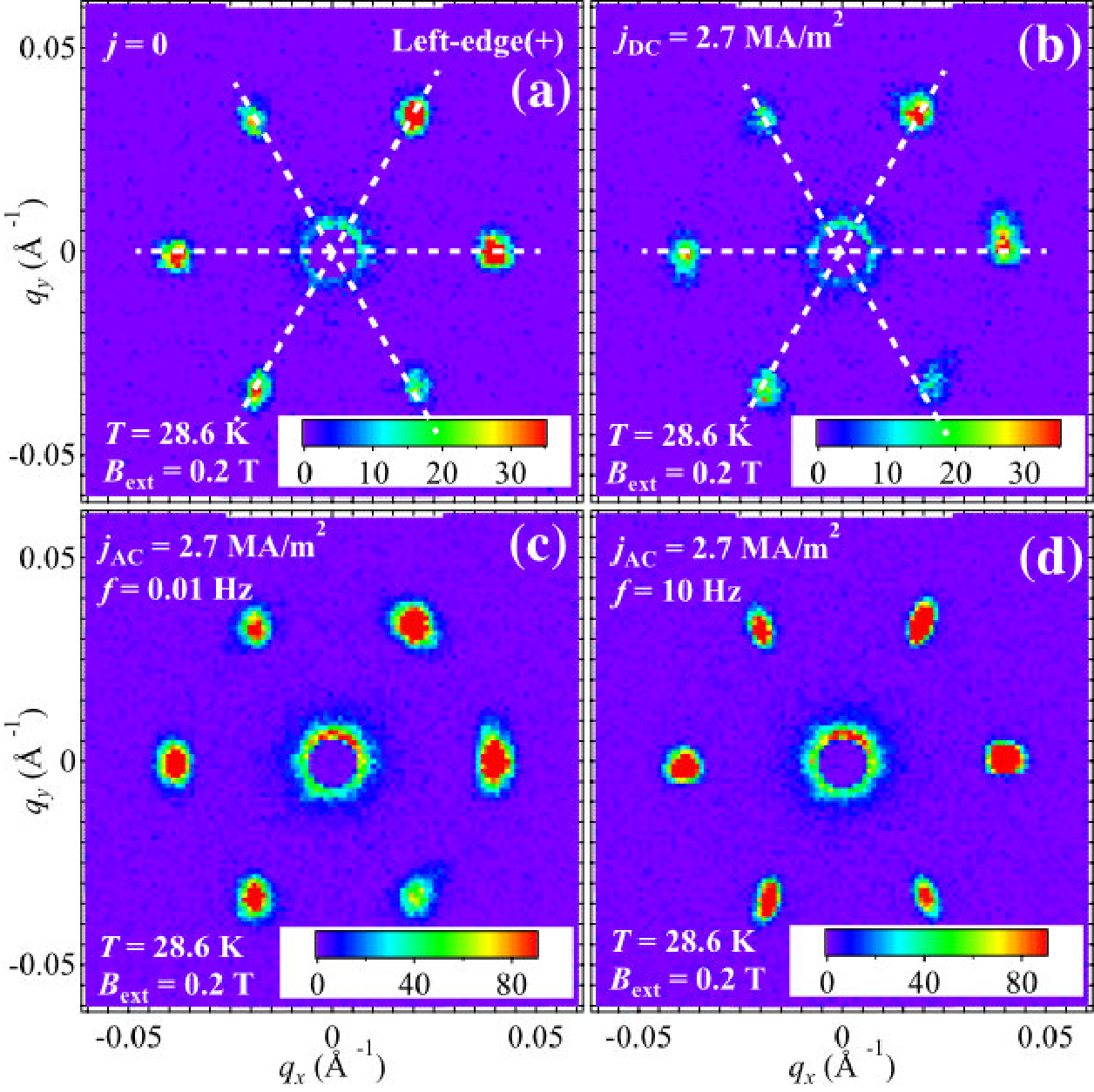}
\end{center}
\caption{\label{fig_08} 
The SANS patterns measured at the left edge (+) of the sample as schematically illustrated in Fig.~\ref{fig_01}(b) (bottom) are shown.  
The measurements were conducted at the pristine state (a), $j_{\mathrm{DC}}$ = 2.7~MA/m$^{2}$ (b), 
and $j_{\mathrm{AC}}$ = 2.7~MA/m$^{2}$ for $f$ = 0.01 (c), 10~Hz (d), respectively.  
All measurements were conducted at $T$ = 28.6~K and $B_{\mathrm{ext}}$ = 0.2~T.  
The data for (a) and (b) were measured for 20~minutes.  
The white dotted lines in (a) and (b) are guides to the eye for the peak positions of the skyrmion reflections in pristine condition.  
For the data (c) and (d), a 60-min measurement was conducted and the AC flow was applied.  
}
\end{figure}

\begin{figure*}
\begin{center}
\includegraphics*[width=160mm,clip]{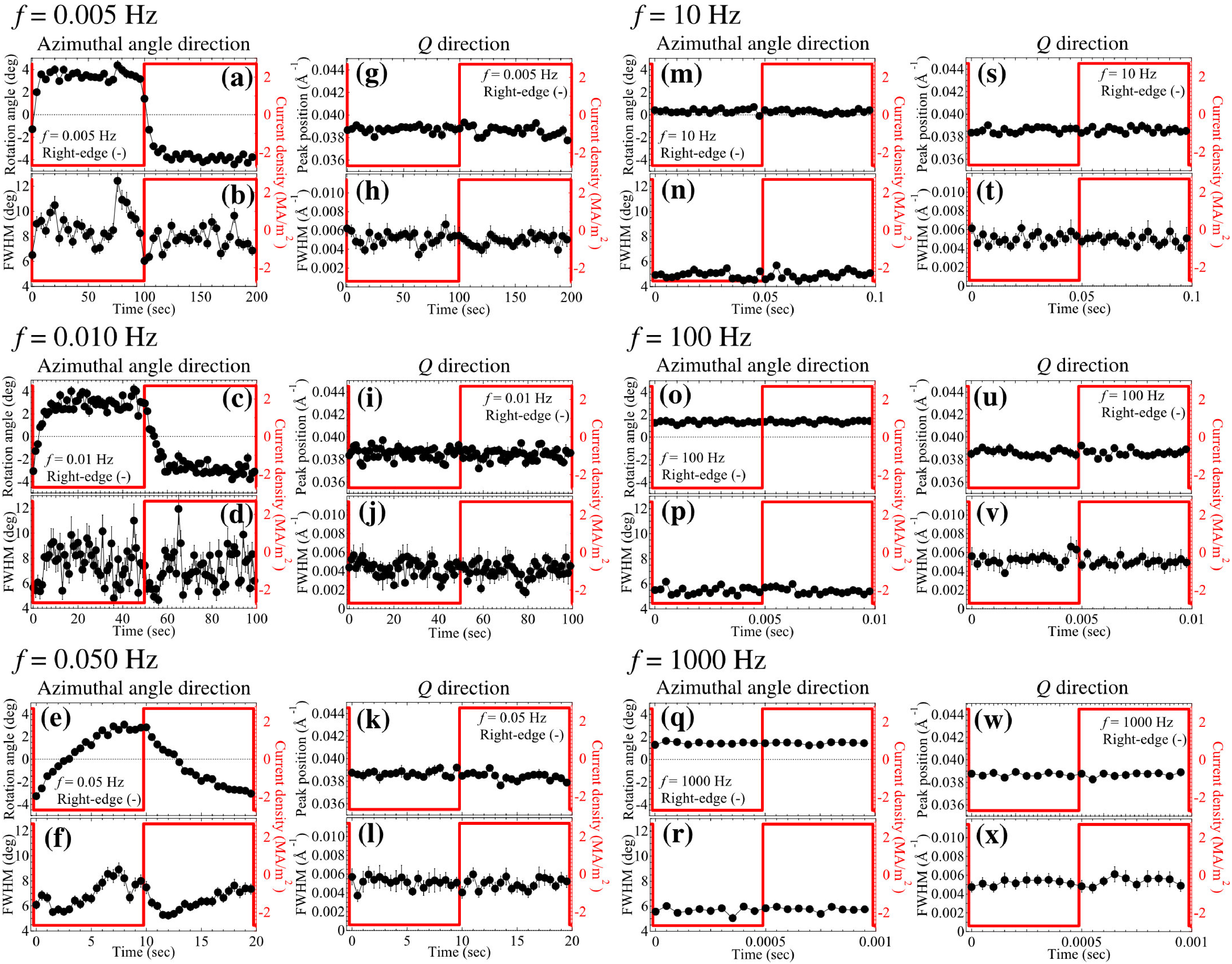}
\end{center}
\caption{\label{fig_09} 
Time dependence of the averaged rotation angle (a,~c,~e,~m,~o,~q) and spot width (b,~d,~f,~n,~p,~r) in the azimuthal angle direction 
of the skyrmion reflections measured at $T$ = 28.6~K and $B_{\mathrm{ext}}$ = 0.2~T at right edge.  
Time dependence of the peak position (g,~i,~k,~s,~u,~w) and spot width (h,~j,~l,~t,~v,~x) to $Q \hat{e}_{Q}$ direction 
of the skyrmion reflections at right-edge.  
The data were measured under the AC with the square wave form at $f$ = 0.005~Hz (a,~b,~g,~h), 0.010~Hz (c,~d,~i,~j), 0.050~Hz (e,~f,~k,~l), 
10~Hz (m,~n,~s,~t), 100~Hz (o,~p,~u,~v), and 1000~Hz (q,~r,~w,~x), respectively.  
Red lines stand for the periodicity and wave form of the AC density.  
}
\end{figure*}

\begin{figure*}
\begin{center}
\includegraphics*[width=160mm,clip]{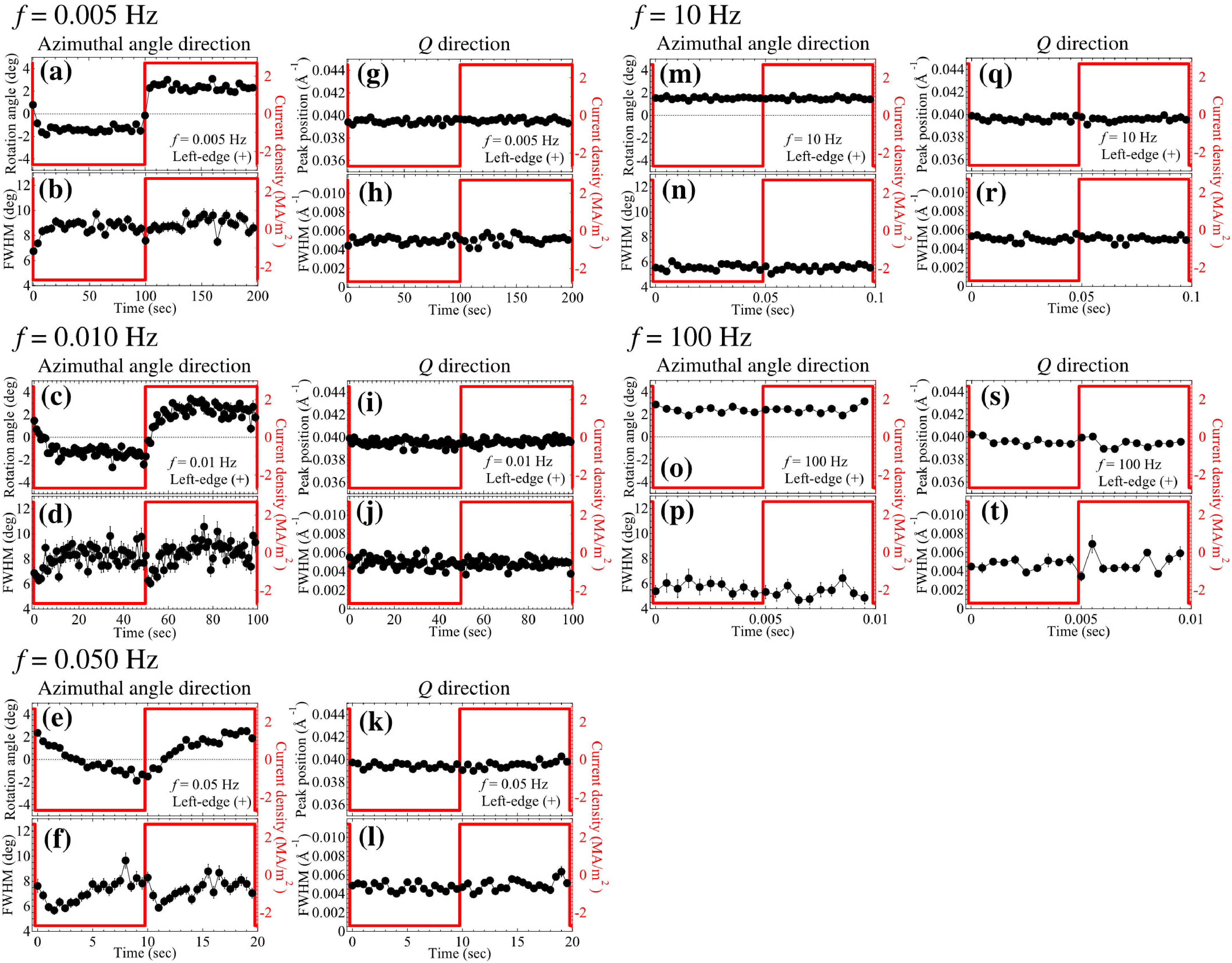}
\end{center}
\caption{\label{fig_10} 
Time dependence of the averaged rotation angle (a,~c,~e,~m,~o) and spot width (b,~d,~f,~n,~p) in the azimuthal angle direction 
of the skyrmion reflections measured at $T$ = 28.6~K and $B_{\mathrm{ext}}$ = 0.2~T at left edge.  
Time dependence of the peak position (g,~i,~k,~q,~s) and spot width (h,~j,~l,~r,~t) to $Q \hat{e}_{Q}$ direction 
of the skyrmion reflections at left-edge.  
The data were measured under the AC with the square wave form at $f$ = 0.005~Hz (a,~b,~g,~h), 0.010~Hz (c,~d,~i,~j), 0.050~Hz (e,~f,~k,~l), 
10~Hz (m,~n,~q,~r), and 100~Hz (o,~p,~s,~t), respectively.  
Red lines stand for the periodicity and wave form of the AC density.  
}
\end{figure*}

Figure~\ref{fig_08} shows the representative time-integrated SANS patterns measured at the left edge.  
In Figs.~\ref{fig_09}~and~\ref{fig_10}, the obtained time dependences of the center position of the rotation angle and the spot width 
in the azimuthal angle direction, and the $Q$ position and the spot width to $Q \hat{e}_{Q}$ direction for each current density 
measured at right and left edges are shown, respectively.  
The center positions of the rotation angle measured at 100 and 1000~Hz at right edge, and 10 and 100~Hz at left edge 
are at $\sim$2$^{\circ}$.  
The possible origin of this shift is the misorientation of the electric current direction at each edge.  
To confirm it, the further experiments with changing the electric current orientation are necessary.  

%\bibliography{mybibfile}

\end{document}